\documentclass[aps,prl,superscriptaddress,amsmath,amssymb,notitlepage]{revtex4-1}
\usepackage{graphicx}
\usepackage{natbib}
\usepackage{color}
\usepackage[dvipsnames]{xcolor}

\newcommand{\beginsupplement}{%
        \setcounter{table}{0}
        \renewcommand{\thetable}{S\arabic{table}}%
        \setcounter{figure}{0}
        \renewcommand{\thefigure}{S\arabic{figure}}%
        \setcounter{page}{1}
     }
     
\definecolor{dgreen}{rgb}{0,0.5,0}
\newcommand{\fg}[1]{\textcolor{black}{#1}} % modifications de Francois, a valider par Sham : les mettre en \modif
\newcommand{\modif}[1]{\textcolor{black}{#1}}  % modifications validees dans le texte principal, pour la version en noir et blanc

\begin{document}

%\preprint{}

%\title{\modif{In a Stokes experiment an epithelial monolayer migrates as a Maxwell viscoelastic fluid}}

%\title{\modif{Interplay between cell deformations and rearrangements in an epithelial monolayer heterogeneous flow. }}

%\title{\modif{Due to interplay between cell deformations and rearrangements, an epithelial monolayer migrates as a Maxwell viscoelastic fluid. }}

\title{\modif{A migrating epithelial monolayer flows like a Maxwell viscoelastic liquid}}



\author{S. Tlili}
 \email{\modif{sham.tlili@univ-amu.fr}}
\thanks{\modif{current address: Institut de Biologie du D\'eveloppement de Marseille, UMR 7288, Case 907, Parc Scientifique de Luminy, 
13288 Marseille Cedex 9, France}}
\affiliation{Laboratoire Mati\`ere et Syst\`emes Complexes, Universit\'e \modif{de Paris - Diderot}, CNRS UMR 7057, 10 rue Alice Domon et L\'eonie Duquet, F-75205 Paris Cedex 13, France}
 \affiliation{Mechanobiology Institute, Department of Biological Sciences, National University of Singapore, 5A Engineering Drive, 1, 117411 Singapore}

\author{\modif{M. Durande}}%
\affiliation{Laboratoire Mati\`ere et Syst\`emes Complexes, Universit\'e \modif{de Paris - Diderot}, CNRS UMR 7057, 10 rue Alice Domon et L\'eonie Duquet, F-75205 Paris Cedex 13, France}
 
\author{C. Gay}%
\affiliation{Laboratoire Mati\`ere et Syst\`emes Complexes, Universit\'e \modif{de Paris - Diderot}, CNRS UMR 7057, 10 rue Alice Domon et L\'eonie Duquet, F-75205 Paris Cedex 13, France}

\author{B. Ladoux}%
\affiliation{Mechanobiology Institute, Department of Biological Sciences, National University of Singapore, 5A Engineering Drive, 1, 117411 Singapore}
 \affiliation{Institut Jacques Monod, Universit\'e \modif{de Paris - Diderot}, CNRS UMR 7592, 15 rue H\'el\`ene Brion, 75205 Paris Cedex 13, France}%

\author{F. Graner}%
\affiliation{Laboratoire Mati\`ere et Syst\`emes Complexes, Universit\'e \modif{de Paris - Diderot}, CNRS UMR 7057, 10 rue Alice Domon et L\'eonie Duquet, F-75205 Paris Cedex 13, France}%

\author{H. Delano\"e-Ayari}%
\email{\modif{helene.delanoe-ayari@univ-lyon1.fr}}
\affiliation{Univ. Lyon, Universit\'e Claude Bernard Lyon 1, CNRS UMR 5306,  Institut Lumi\`ere Mati\`ere, Campus LyonTech - La Doua,  Kastler building, 10 rue Ada Byron,  
F-69622 Villeurbanne Cedex, France}%

\date{\today}

\begin{abstract}
 We perform a bidimensional Stokes experiment in an active cellular material: \modif{an autonomously migrating} monolayer of Madin-Darby Canine Kidney (MDCK) epithelial cells \modif{flows} around a circular obstacle within a long and narrow channel, involving \modif{an interplay between} cell shape changes and neighbour rearrangements. Based on image analysis of tissue flow and \modif{coarse-grained cell anisotropy}, we determine the tissue strain rate, cell deformation and rearrangement rate fields, which are spatially heterogeneous. We find that the cell deformation and rearrangement rate fields correlate strongly, which is compatible with a \modif{Maxwell} viscoelastic liquid behaviour \modif{(and not with a Kelvin-Voigt viscoelastic solid behaviour)}. \fg{ The value of the associated} relaxation time
\fg{is measured as}  $\tau = 70 \pm 15$~min, \fg{is observed to be independent of}  \modif{obstacle size and division rate, and \fg{is} increased by inhibiting myosin activity}. In this experiment, the monolayer behaves as a flowing material with a Weissenberg number close to one which shows that both elastic and viscous effects can have comparable contributions in the process of collective cell migration.
\end{abstract}

%The migration around the obstacle can generate strain rates of $5\,10^{-3}$~min$^{-1}$  and the viscoelastic relaxation time, %independent on the obstacle size, is $\tau = 70$$\pm$15~min. We show that the monolayer behaves as a flowing material with a %Weissenberg number close to one where both elastic and viscous effects have comparable contributions.
%\keywords{}

\maketitle

%\subsection{Introduction}

%\paragraph{\textbf{Introduction}}

 Epithelial tissues are active cellular materials made of constitutive objects, the cells, that can not only deform and exchange neighbors, but also grow, divide, have a polarity and exert active stresses \cite{Zehnder2015,Puliafito2012,LADOUX2016420,C5SM01382H}. 
Tissue mechanical properties have a crucial importance in biological morphogenetic processes \cite{Heisenberg2013}. In particular, cell neighbour rearrangements contribute to shaping tissues, as during the Drosophila embryo germ band extension \cite{Collinet2015}. Rearrangements are often described as either passive, caused by an external stress emerging at the tissue scale; or active, for instance triggered and directed by an anisotropic distribution of molecules (such as myosin or cadherins) at the cell-cell contacts \cite{Guirao2017}.  Active cell contour fluctuations are key ingredients to fluidize tissues \modif{so that they flow under tissue-scale stress,} both \textit{in vitro}  \cite{Marmottant2009} and \textit{in vivo} \cite{Mongera2018}.

Experiments performed on embryonic tissues \cite{Serwane2017}, multicellular spheroids \cite{Marmottant2009,Guevorkian2010}, or cell monolayers with or without substrate \cite{Vincent2015,Harris2012} have suggested describing  tissues as \modif{viscoelastic liquids}. However, there is still a debate around the microscopic origin and value of the viscoelastic relaxation time $\tau$ and on whether MDCK monolayers behave predominantly as liquids or solids \cite{Vincent2015,Tlili2018}. MDCK  cells can actively migrate on a substrate while sustaining tissue cohesivity \modif{\cite{Reffay2011,Serra-Picamal2012,Harris2012,Doxzen2013,Cochet2013}}. Collective cell migration is due to the cells displaying simultaneously cryptic lamellipodia on their basal side and adherent junctions on their apical side \cite{Farooqui51,Das2015}. Studying their bidimensional flow   in long and narrow adhesive strips  facilitates experiments, simulations, theory and their mutual comparisons \modif{\cite{Vedula2012}}.  

The flow of a liquid relative to a circular obstacle, whether in two or three dimensions, is a classical experiment \modif{(}introduced by Stokes \cite{Stokes1850}\modif{)} to measure a viscosity, and/or to probe the mechanical behaviour of viscoelastic or viscoplastic materials \cite{Saramito2016}. More complex materials which are together viscous, elastic and plastic such as liquid foams  (bubbles are able to sustain elastic deformations due to their surface tension, until a yield point where they rearrange and exchange neighbours \cite{Cantat2013,Weaire1999}) 
have been well characterised using the Stokes experiment\modif{, in which upstream bubbles are elongated tangentially to the obstacle and downstream bubbles are elongated radially} \cite{Cheddadi2011}.
 The Stokes flow geometry has even been applied in contexts as different as granular materials \cite{PhysRevE.87.032207}, or active materials:  to study clogging in animal crowds \cite{Zurigel2014}, or cell-substrate interaction  in  MDCK cell monolayers, where \modif{average traction forces were found to pull the monolayer} towards the obstacle \cite{Kim2013}.

The total deformation rate is the strain rate, \modif{also called velocity gradient}, $\dot{\varepsilon}_{tot}$.
It is the sum of the \modif{coarse-grained cell} deformation rate and of the intercellular topological change rate \modif{(hereafter ``rearrangement rate" for short)}:  \modif{$\dot{\varepsilon}_{tot}=\dot{\varepsilon}_{cell}+\dot{\varepsilon}_{r}$}  \cite{Blanchard2009}\fg{. The former} is the time derivative of the \modif{coarse-grained  cell deformation field $\varepsilon_{cell}$},
\fg{while in the} absence of cell divisions and apoptoses\fg{, the latter} reduces to the  \modif{deformation rate due to cell-cell rearrangements, $\dot{\varepsilon}_{r}$}.
\modif{T}he Stokes flow, \modif{in which} these fields are heterogeneous\modif{,} displays simultaneously a variety of combinations of these fields, depending on the position with respect to the obstacle; for instance there are \modif{regions} where the strain rate and deformation are parallel, other \modif{regions} where they are orthogonal.  In that respect, to discriminate between different models, the Stokes flow geometry is more efficient  \cite{Cheddadi2011} than a homogeneous flow such as \modif{a pure shear} between parallel \modif{boundaries} (Couette flow) \cite{Cheddadi2012}. 

\begin{figure}[ht]
\centering
\includegraphics[width=17.9cm]{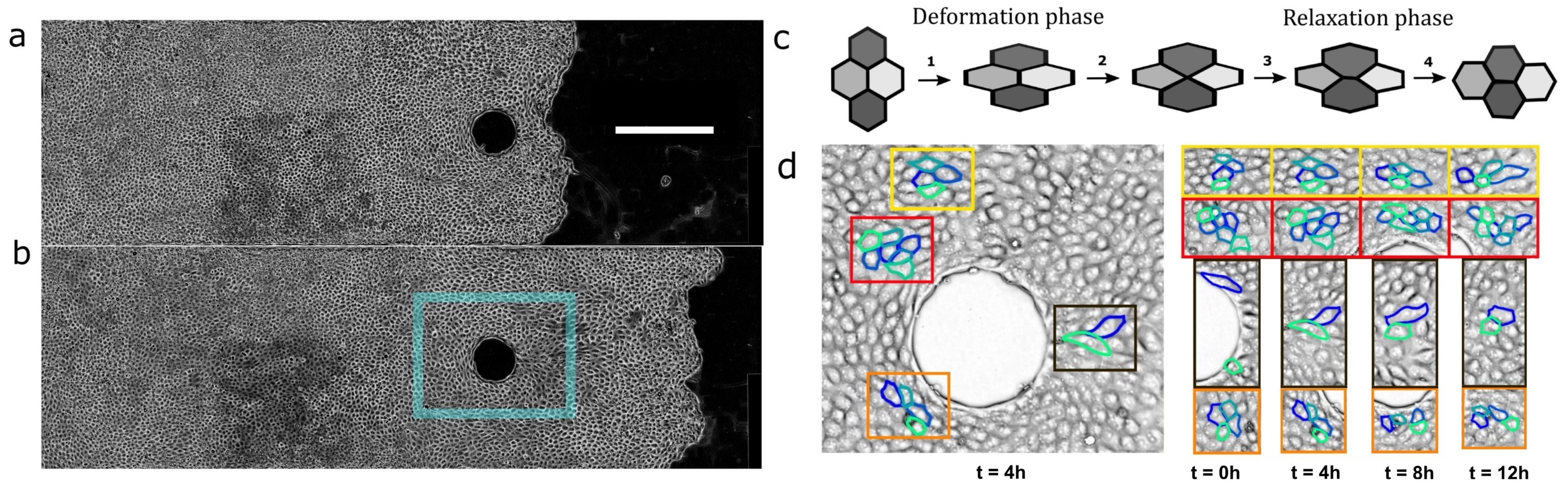}

\caption{Stokes flow of migrating cells. Initial time $t=0$  $(a)$, and  $t=12$ hours later $(b)$, phase contrast images  of a monolayer migrating from left to right around a 200~$\mu$m diameter obstacle (see Supplementary Movie 1); strip width 1000~$\mu$m, length 4~mm (the right part is longer than visible on the pictures), scale bar 500~$\mu$m. 
$(c)$ Sketch of \modif{cell deformation and cell-cell rearrangement} driven by tissue velocity gradient: (1) cells deform and \modif{a} cell-cell junction shortens; (2) \modif{the} cell-cell junction shrinks, and a four-fold vertex is formed; (3) a new pair of neighbour cells is formed \modif{(rearrangement)}; (4) its junction lengthens and cell\modif{s deform again, during their shape relaxation}.
$(d)$ Four zones, highlighted by color frames (left\modif{, $t=4$~h}), are tracked on $t=0$, 4, 8 and 12~h (right), with the same color code, to evidence a few examples of cell\modif{-cell} rearrangements. }

\label{fig:exp_base}
\end{figure}

\begin{figure}[ht]
%\centering
%$(a)$ \hfill $(b)$ \hfill $(c)$ \hfill ~ \\
 % \includegraphics[width=5.9cm]{pivfield.jpg}
 % \includegraphics[width=5.9cm]{strainratebars.jpg}
 % \includegraphics[width=5.9cm]{deformationbars.jpg}
  \includegraphics[width=17.9cm]{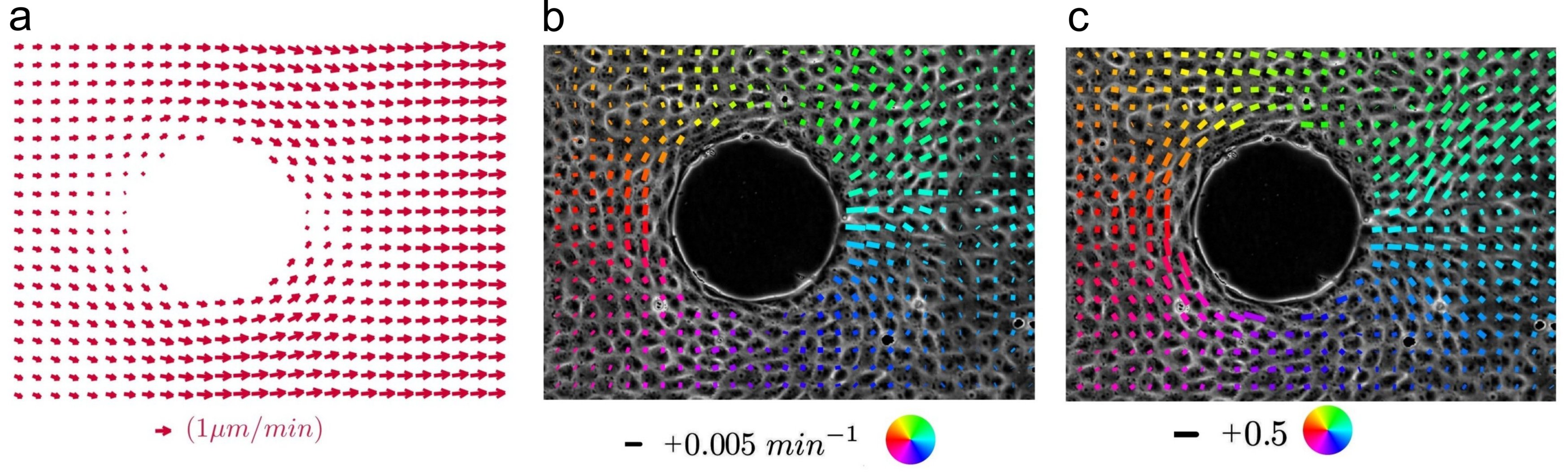}
\caption{Cell velocity and deformation field maps around the obstacle. $(a)$ Velocity  field $\vec{v}(x,y)$ averaged over 8 hours. 
$(b)$ Deviator of the velocity gradient tensor \modif{symmetric part,} $\nabla v_{\rm sym}^{\rm dev}$\modif{,} obtained by spatial derivation of the velocity field in (a). It is diagonalized and each bar represents its main axis of extension. $(c)$ \modif{Deformation} tensor deviator $\mathbf{\varepsilon_{\rm e}^{\rm dev}}$. It is diagonalized and each bar represents its main axis of extension. Same \modif{experiment} as in Fig. \ref{fig:exp_base}, obstacle diameter 200~$\mu$m; scales are indicated below each panel. In (b,c) the color codes for the \modif{angular position of each point, in polar coordinates originating at the obstacle center.}}
\label{fig:kinematic}
\end{figure}

\begin{figure}[ht]
%\centering
%$(a)$ \hfill $(b)$ \hfill $(c)$ \hfill ~ \\
  %\includegraphics[width=5.9cm]{diffsizefit.jpg}
  %\includegraphics[width=5.9cm]{oneobstaclefit.jpg}
 % \includegraphics[width=5.9cm]{overlapdpe1.jpg}
\includegraphics[width=17.9cm]{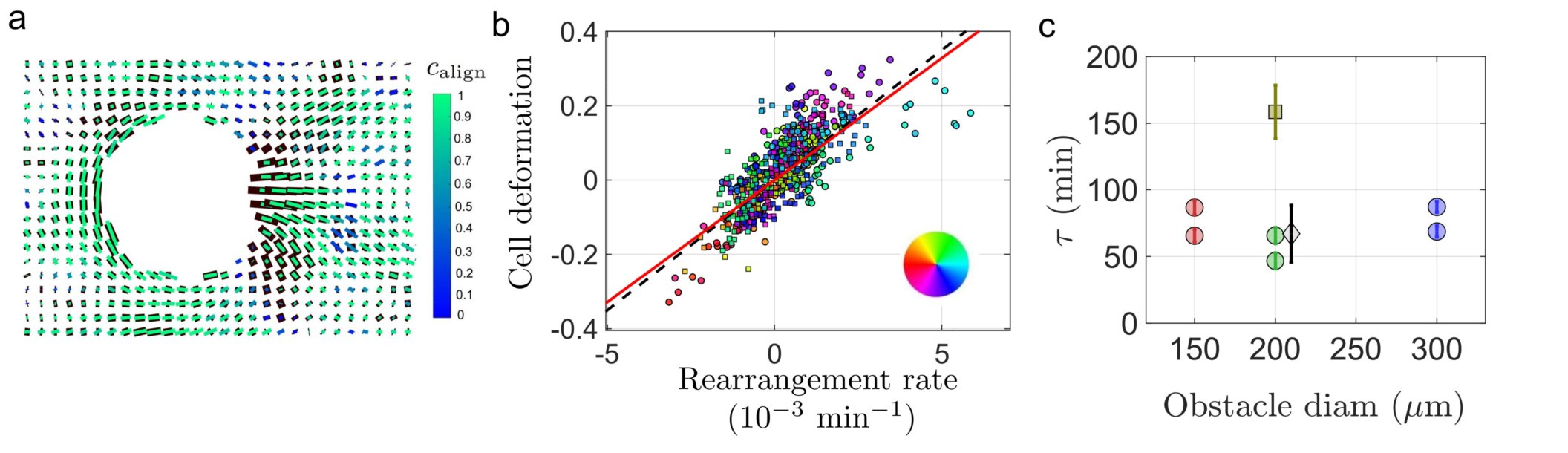} 
\caption{ 
\modif{Test of Maxwell  model and measurement of viscoelastic time.
$(a)$ 
Cell deformation vs rearrangement rate: correlation map corresponding to the same individual experiment as in  Figs. \ref{fig:exp_base}, \ref{fig:kinematic}; obstacle diameter 200~$\mu$m.
In black: deviatoric part of the rearrangement rate \modif{$\dot{\varepsilon}_{r}$}, multiplied by \modif{$\tau = 70$~min}.
In color, deviatoric part of the cell deformation $\varepsilon_{\rm e}$. The color codes for the alignement coefficient $c_{\rm align}$ of these tensors, namely the square cosine of their relative angle, with green corresponding to tensors being aligned ($c_{\rm align} =1$) and blue to  tensors being orthogonal ($c_{\rm align} =0$).
$(b)$ 
Cell deformation vs rearrangement rate: components of the tensor deviatoric parts. Circles: $ \langle\left(\varepsilon_{\rm e}\right)_{xx}-\left(\varepsilon_{\rm e}\right)_{yy}\rangle/2$ plotted vs  $ \langle\left(\dot{\varepsilon}_{r} \right)_{xx}-\left(\dot{\varepsilon}_{r}\right)_{yy}\rangle/2$. Squares: $\langle\left(\varepsilon_{\rm e}\right)_{xy}\rangle$
 plotted vs $\langle\left(\dot{\varepsilon}_{r}\right)_{xy}\rangle$. 
Same experiment as in (a), color coding for the box position angle as in Fig. \ref{fig:kinematic}b,c. 
Solid \modif{red line: linear fit to the data, passing through the origin, slope 68~min; dashed black line:} slope 70~min. 
%150 microns 
%1 : 0.64
%2 : 0.77
%
%200 microns 
%3 : 0.68
%4 : 0.53
%
%300 microns  
%5 : 0.76
%6 : 0.63
$(c)$ 
Circles: values of $\tau$ found for the six strips \modif{(Fig.~\ref{fig:supp3obstacles}): namely} two 150~$\mu$m obstacles (red), two 200~$\mu$m obstacles (green) and two 300~$\mu$m obstacles (blue). Square: with blebbistatin (Fig.~\ref{fig:suppdrugs}a,b). Diamond: without mitomycin (Fig.~\ref{fig:suppdrugs}c,d)}\modif{, slightly shifted to the right for legibility.} 
 }
\label{fig:superimpos}
\end{figure}

Here, we observe  over \modif{up to one day} the Stokes flow of MDCK cell monolayers (Fig. \ref{fig:exp_base}a,b) \modif{with many} cell rearrangements (Fig. \ref{fig:exp_base}c,d)\modif{, and with only few cell divisions thanks to a drug}.
\modif{Direct} image analysis methods yield velocity, velocity gradient, and \modif{coarse-grained cell anisotropy} fields \modif{within the frame of continuum mechanics} (Fig. \ref{fig:kinematic}).
We deduce the rearrangement rate field and quantify the field correlations to probe the tissue rheology (Fig. \ref{fig:superimpos}).

%\onecolumngrid % a remettre si on veut remettre le papier en \twocolumn

%\twocolumngrid % a remettre si on veut remettre le papier en \twocolumn

%\paragraph{\textbf{Methods}}
The procedure for micropattern printing,  cell culture, imaging and velocity measurement is described in details in Ref.  \cite{Tlili2018}. 
Briefly, a strip is 4~mm long; it is adhesive, while its four boundaries and the circular obstacle are not.
To check reproducibility,  each experimental batch is composed of two identical strips. To test the effect of experiment dimensions, three batches are used: a first one with obstacle diameter 150~$\mu$m and strip width 750~$\mu$m; a second one with obstacle diameter 200~$\mu$m and strip width 1000~$\mu$m; a third one with obstacle diameter 300~$\mu$m and strip width 1000~$\mu$m. 

We inhibit cell divisions using mitomycin; this prevents cell density increase and jamming that slows down migration. \modif{Two hours after having added the mitomycin, we take the first image of the movie and define it as $t = 0$.} 
MDCK monolayers can then migrate \modif{over millimeters and during  \modif{many hours} at constant \modif{total} cell number}. 
The 2 mm long region upstream of the obstacle serves as a cell reservoir; its \modif{two-dimensional} density is initially high\fg{, then} decreases.
The typical time for cells to migrate over a distance equivalent to the obstacle diameter of 200~$\mu$m is 3 h.  For obstacle diameters of 400~$\mu$m or larger, \modif{the transit time of a cell to migrate across the whole zone influenced by the obstacle is comparable to the experiment duration, so that the steady flow is not completely established.}
For obstacle diameters of  100~$\mu$m  or smaller, cells \modif{often}  \fg{form} suspended bridges over the non-adhesive obstacle.

%\onecolumngrid % a remettre si on veut remettre le papier en \twocolumn

%\twocolumngrid % a remettre si on veut remettre le papier en \twocolumn

\modif{To measure the velocity field and total deformation rate, w}e divide the field of view in \modif{square boxes} of 128 pixel side \modif{(83.2~$\mu$m), labeled by their position $(x,y)$}, containing typically 10 cells. We measure the two-dimensional velocity field  $\vec{v}(x,y,t)$  using a custom-made Matlab optic flow code based on the Kanade Lucas Tomasi (KLT) algorithm \cite{Lucas1981} with a level 2 pyramid. 
This coarse-grained  velocity field can be averaged in time (see Fig. \ref{fig:kinematic}a), yielding components $v_i(x_j)$ where $i,j=1$ or 2.
Using finite differences we obtain in each box the symmetric part of the velocity gradient, $\modif{\dot{\varepsilon}_{tot} = }\nabla \vec{v}_{\rm sym}$, with components $(\partial_i v_j + \partial_jv_i)/2$. This symmetric tensor can be diagonalized\modif{. Its anisotropic part is its deviator $\dot{\varepsilon}_{tot}^{\rm dev}= \dot{\varepsilon}_{tot}- {\rm Tr}(\dot{\varepsilon}_{tot})I/2$, where $I$ is the unit tensor in 2 dimensions; $\dot{\varepsilon}_{tot}^{\rm dev}$ has} two opposite eigenvalues. We graphically represent in each \modif{box} (Fig. \ref{fig:kinematic}b) \modif{this} anisotropic part,  with bars oriented along the direction of the eigenvector corresponding to the positive eigenvalue (local tissue expansion direction) and of length proportional to \modif{this} eigenvalue. % perpendicular direction carries the same information, with the opposite eigenvalue, and it is unnecessary to plot it.

\begin{figure}[ht]
%\centering
%$(a)$ \hfill $(b)$ \hfill $(c)$ \hfill ~ \\
 % \includegraphics[width=5.9cm]{zoomobstaclees.jpg}
 % \includegraphics[width=5.9cm]{fourierellipses.jpg}
 % \includegraphics[width=5.9cm]{cellellipses.jpg}
   \includegraphics[width=17.9cm]{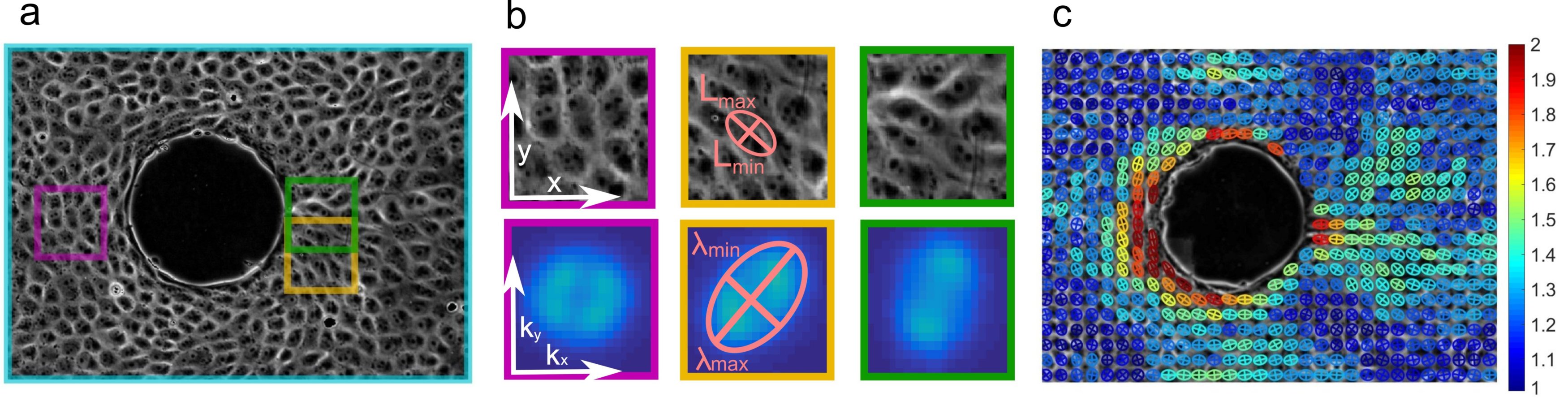}
\caption{Fourier Transform. $(a)$ Phase contrast image of the monolayer, same strip as in Fig. \ref{fig:exp_base}.
$(b)$ 2D Fourier analysis for three different examples of local cell patterns identified in purple, yellow and green \modif{box}es in (a). The top panels are in real space and the bottom panels are in Fourier space, with axes indicated in the purple \modif{box}es (left). Each Fourier image is blurred using a Gaussian filter and averaged over 8 h. The inertia matrix of the 5$\%$ brightest pixels of the image is  diagonalized and the pattern is represented as an ellipse (middle, bottom) of axes  $\lambda_{max}$, $\lambda_{min}$, then their inverses build the ellipse in real space  with axes $L_{max}$, $L_{min}$ (middle, top). $(c)$ Fourier transform map; colors code for the \modif{coarse-grained} cell anisotropy $L_{max}/L_{min}$ from 1 (blue) to 2 (red). }
\label{fig:supp1fourier}
\end{figure}

To extract the 
 \modif{coarse-grained cell anisotropy (i.e. anisotropy of the average cell shape, not to be confused with the average of the cell shape anisotropy)} 
we use Fourier Transform (FT, see Fig. \ref{fig:supp1fourier}); details and validations are provided in \modif{Ref.} \cite{Durande2019}. This method provides an efficient \modif{estimation} of the \modif{coarse-grained elastic deformation
field anisotropy} without having to recognize and segment each individual cell contour \modif{(as in refs. \cite{Guirao2015,Merkel2017})} which is challenging on phase contrast images.
%Even if the pattern is not well contrasted, its FT provides an efficient measurement thanks to time averaging, with sufficient precision for the present study, without having to recognize and segment each individual cell contour, and thus without limitation on the cell number. 
We use the same grid and \modif{box} size as for the velocity. 
In each \modif{box}, we multiply the image by a windowing function to avoid singularities in the FT due to  boundaries. We then compute the FT using the Fast Fourier Transform algorithm implemented in Matlab (\textit{fft2.m})
and keep only its amplitude (not the phase), which can be time averaged to increase signal-to-noise ratio.
%and average it in time over 8 hours. 
We binarize the resulting Fourier space pattern keeping 5\% of the brightest pixels (for justification of this percentile value, see \modif{Ref.} \cite{Durande2019}). We compute the inertia matrix of the binarized FT pattern and diagonalize it, which yields two eigenvalues ${\lambda_{\rm max}^{2}}$ and ${\lambda_{\rm min}^{2}}$ in the directions of the pattern main axes. 
They determine the ellipse back in the real space, with eigenvalues $L_{\rm max}=\frac{2m}{\lambda_{\rm min}}$ and $L_{\rm min}=\frac{2m}{\lambda_{\rm max}}$ in the directions of the same axes with $m$ the size of the FT image in pixels.

We define as follows the \modif{coarse-grained elastic} deformation tensor $\mathbf{\varepsilon_{\rm e}}$ with respect to a rest state which we assume to be isotropic, with two equal eigenvalues $L_0 = \sqrt{L_{\rm max}L_{\rm min}}$: $\mathbf{\varepsilon_{\rm e}}$  has the same eigenvectors as the Fourier pattern, and two eigenvalues $ \frac{1}{2}(\frac{L_{\rm max}^{2}}{L_0^{2}}-1)$ and  $\frac{1}{2}(\frac{L_{\rm min}^{2}}{L_0^{2}}-1)$. 
The deformation tensor $\mathbf{\varepsilon_{\rm e}}$ is equivalent to other definitions of the strain (e.g. true strain \cite{Graner2008,Durande2019}) within a linear approximation; in addition, $\mathbf{\varepsilon_{\rm e}}$ has the advantage to have well established transport equations \cite{Tlili2015}. 
%We choose this definition $\mathbf{\varepsilon_{\rm e}}$ because within a linear approximation  it is equivalent to other usual definitions, including the true strain \cite{Graner2008,Durande2019} but with this definition, $\mathbf{\varepsilon_{\rm e}}$ has well established transport equations \cite{Tlili2015}. 
%$\mathbf{\varepsilon_{\rm e}}$ is a linear approximation of the true elastic strain \cite{Graner2008,Durande2019} and  has well established transport equations \cite{Tlili2015}.
 We represent the anisotropic part of  $\mathbf{\varepsilon_{\rm e}}$, namely its deviator $\mathbf{\varepsilon_{\rm e}^{\rm dev}}$, which only depends on the ratio $\frac{L_{\rm max}}{L_{\rm min}}$. This represents the 
 \modif{coarse-grained elastic deformation field} anisotropy
depicted  as a bar in Fig. \ref{fig:kinematic}c.  %\modif{We have ${\rm Tr}(\dot{\varepsilon}_{tot})  = {\rm Tr}(\dot{\varepsilon}_{cell})+ {\rm Tr}(\dot{\varepsilon}_{r})$, $\dot{\varepsilon}_{tot}^{\rm dev} =\dot{\varepsilon}_{cell}^{\rm dev}+\dot{\varepsilon}_{r}^{\rm dev}$;  and the anisotropic part of cell deformation $\varepsilon_{cell}^{\rm dev}$ is none other than $\varepsilon_{e}$. 
\fg{In summary, $\varepsilon_{e}^{\rm dev}$ can easily be measured, and it correctly approximates
 $\varepsilon_{cell}^{\rm dev}$.
}
 
\modif{
Since divisions are inhibited, ${\rm Tr}(\dot{\varepsilon}_{r})$ is negligible compared with $\dot{\varepsilon}_{r}^{\rm dev}$ \cite{Guirao2015}; we have ${\rm Tr}(\nabla \vec{v}_{\rm sym}) \approx {\rm Tr}(\dot{\varepsilon}_{cell})$ and $\nabla \vec{v}_{\rm sym}^{\rm dev} =\dot{\varepsilon}_{cell}^{\rm dev}+\dot{\varepsilon}_{r}^{\rm dev}$. Since $\varepsilon_{cell}^{\rm dev} 
\approx
\varepsilon_{e}^{\rm dev}$, we deduce the time averaged rearrangement rate $\langle\dot{\varepsilon}_{r}^{\rm dev} \rangle$ by measuring the difference $\langle \nabla \vec{v}_{\rm sym}^{\rm dev}-\dot{\varepsilon}_{\rm e}^{\rm dev} \rangle$  as in  \cite{Blanchard2009}.
}
From the measurement of $\varepsilon_{\rm e}$, we estimate $\dot{\varepsilon}_{\rm e}$ by taking  into account the cell deformation advection in the flow, as $\dot{\varepsilon}_{\rm e} = \frac{D\varepsilon_{\rm e}}{D t} =
\frac{\partial \varepsilon_{\rm e}}{\partial t}+\vec{v}\cdot \nabla \varepsilon_{\rm e}$: this significantly improves the results presented below  (we still neglect rotation terms \cite{Graner2008,Tlili2015}, which do not change the present results, see \modif{Supplementary Materials and} Fig. \ref{fig:supp2transport}). We correlate the deviator of the cell deformation $\langle \varepsilon_{\rm e}^{\rm dev}\rangle$ with the deviator of the rearrangement rate $\langle \dot{\varepsilon_{\rm r}}^{\rm dev}\rangle$ where both fields are time-averaged over the experiment duration (at least 10 hours).

%\onecolumngrid % a remettre si on veut remettre le papier en \twocolumn
%\twocolumngrid % a remettre si on veut remettre le papier en \twocolumn

%\paragraph{\textbf{Results}}

%Qualitatively, visual observation indicates that the fields of velocity (Fig. \ref{fig:kinematic}a), velocity gradient anisotropic part (Fig. \ref{fig:kinematic}b), cell deformation anisotropic part (Fig. \ref{fig:kinematic}c)  and rearrangement rate (Fig. \ref{fig:superimpos}c, black) are smooth: this is compatible with continuum mechanics. It opens the possibility to look for a partial differential equation which would predict these fields. 
 
\modif{Qualitatively, to model the flow, we note that cell deformation and rearrangement rate fields correlate strongly (Fig. \ref{fig:superimpos}a,b). 
This is compatible with a simple Maxwell viscoelastic liquid model where an elastic intracellular \fg{component} is in series with \fg{a viscous} intercellular} \fg{component representing cell rearrangements \cite{Tlili2015}, with the same stress in both elements:} $ E\varepsilon_{\rm e} =  \eta_r \dot{\varepsilon}_{r}^{\rm dev}$\fg{. H}ere $E$ is an \modif{effective} intracellular Young modulus and $\eta_r$ an \modif{effective} intercellular viscosity due to rearrangements\modif{, or equivalently} $\varepsilon_{\rm e}^{\rm dev}=\dot{\varepsilon}_{r}^{\rm dev}\tau$ with $\tau = \eta_r / E$.  

\begin{figure}[ht]
\begin{center} % a enlever si on veut utiliser l'environnement "figure"
\includegraphics[width=17.9cm]{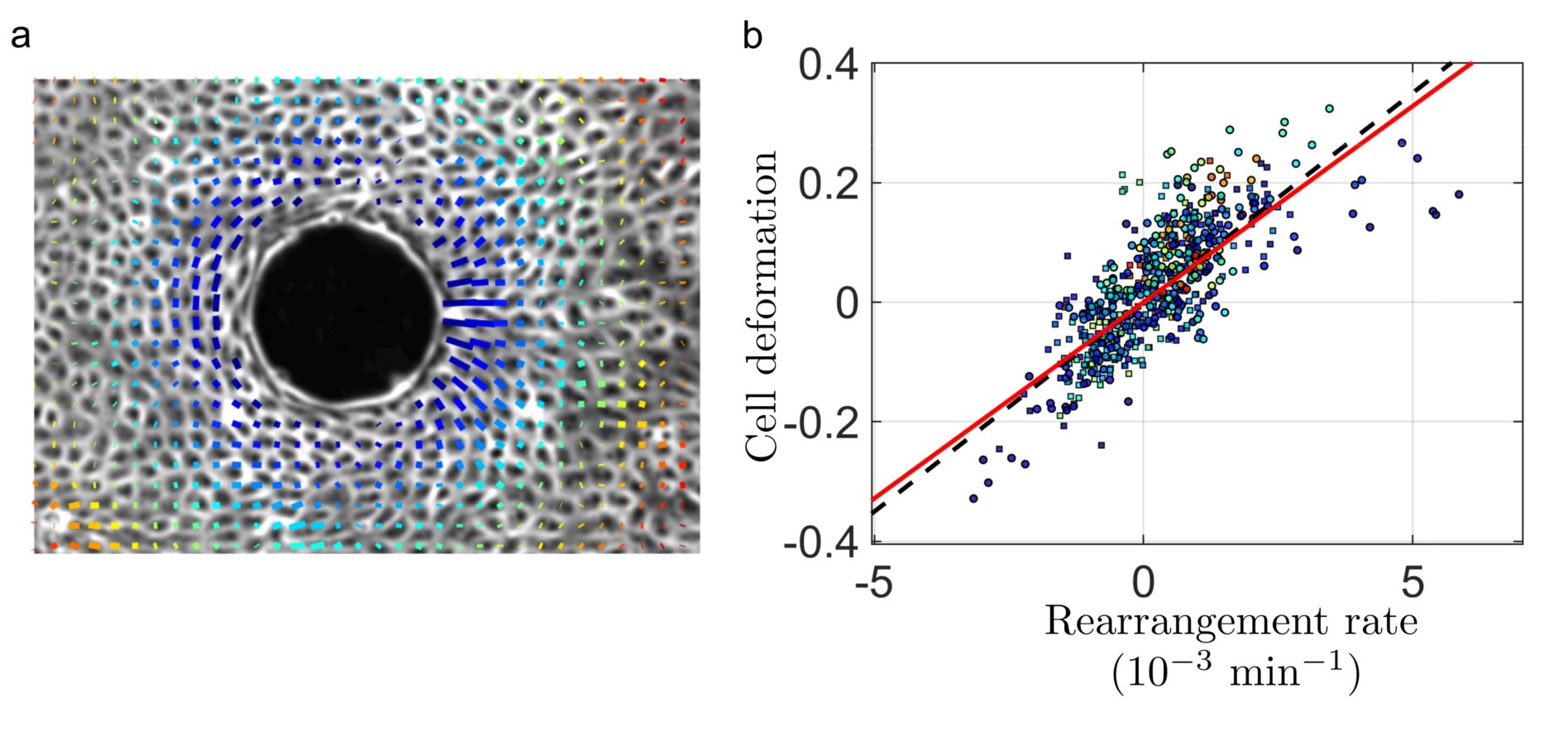}
\end{center} % a enlever si on veut utiliser l'environnement "figure"
\caption{\modif{Range of obstacle effect on flow. Same experiment as in  Fig.~\ref{fig:superimpos}a,b with a different representation. Here, on both panels, the color code of points corresponds to their distance to the obstacle. \modif{For legibility, on the left panel only the cell deformation is represented, not the rearrangement rate.}
}}
\label{fig:distanceplot}
\end{figure}

\modif{To test the Maxwell model, we compare the} anisotropic parts of cell deformation  and rearrangement rate fields. The\modif{ir} orientations are identical, except in a region where both fields are small (right of \modif{the obstacle on} Fig. \ref{fig:superimpos}a, coded in blue). 
\modif{Their amplitudes are proportional to each other (Fig. \ref{fig:superimpos}a}\modif{,b, Fig. \ref{fig:supp3obstacles}).
Their proportionality coefficient is found by linear fit to the data performed with Matlab \textit{robustfit.m}}. 
Each individual experiment provides a self-sufficient data set \modif{with a large enough variation range} thanks to the Stokes flow heterogeneity\modif{. Data with $\varepsilon_{\rm e}^{\rm dev}$ amplitude smaller than 0.05 \fg{correspond to vast, quiet regions of the flow
and} are excluded from the fit \fg{to increase the signal-to-noise ratio}. Points at different distances from the obstacle have similar contributions to the measurement of the viscoelastic time (Fig. \ref{fig:distanceplot}).}

\begin{figure}[ht]
\begin{center} % a enlever si on veut utiliser l'environnement "figure"'''/'';,,,,,,,,,''

\includegraphics[width=9cm]{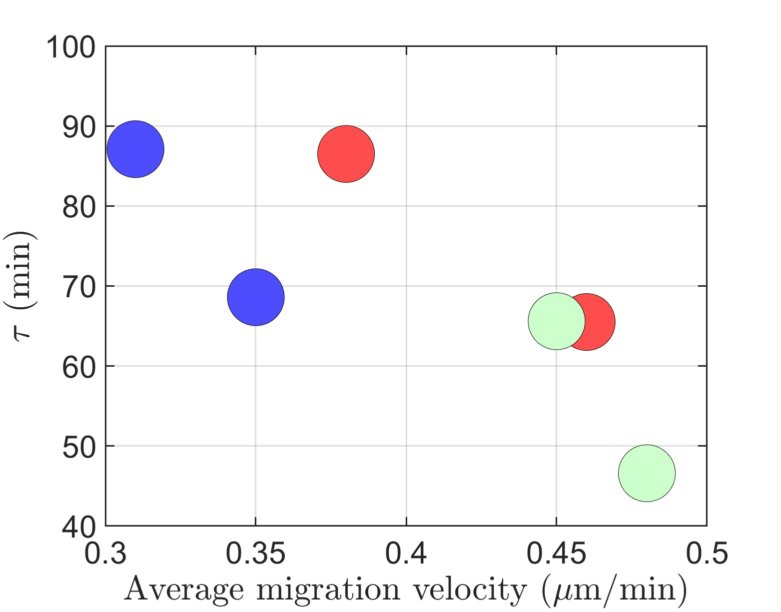}
\end{center} % a enlever si on veut utiliser l'environnement "figure"
\caption{\modif{Viscoelastic time versus monolayer average migration speed, which is the velocity component along $x$ spatially averaged around the obstacle (on the whole field of view) and temporally averaged during the same duration as the one used for time averaging in the analysis. One point per experiment, same color code as for  Fig.~\ref{fig:superimpos}c.}}
\label{fig:rheofluid}
\end{figure}

\modif{Quantitatively, we find the value $\tau = 70\pm15$~min (Fig. \ref{fig:superimpos}a), \fg{whichever} the fit method: 
either by 
fitting the superimposition of all six experiments (Fig. \ref{fig:supp3obstacles}, correlation value $R=0.67$),
or by 
averaging six experimental fits (Fig. \ref{fig:superimpos}c, Fig. \ref{fig:supp3obstacles}, correlations values range from $R=0.53$ to 0.77).
As expected for a characteristic of the material itself, $\tau$ is
 independent on the obstacle dimension (Fig. \ref{fig:superimpos}c).}
\fg{We also plot $\tau$ versus the monolayer average migration velocity around the obstacle and find that it decreases from around 90~min to below 60~min (Fig. \ref{fig:rheofluid}). If this average velocity is used as a proxy of the deviatoric deformation rate, this suggests that the monolayer rheo-fluidifies.}

%\paragraph{\textbf{Discussion}}

\begin{figure}[ht]
\begin{center} % a enlever si on veut utiliser l'environnement "figure"
\includegraphics[width=9cm]{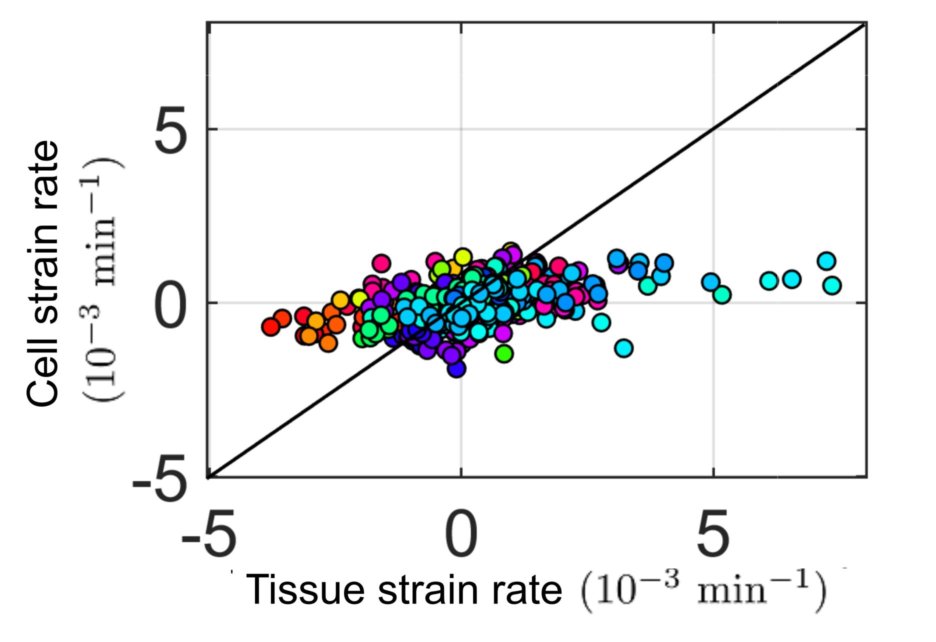}
\end{center} % a enlever si on veut utiliser l'environnement "figure"
\caption{\modif{ Rejection of the Kelvin-Voigt model. 
Plot of the time averaged cellular strain rate  $\langle\frac{\partial \varepsilon_{\rm e}}{\partial t}+\vec{v}\cdot \nabla \varepsilon_{\rm e} \rangle $ versus the tissue strain rate $\langle \nabla v_{\rm sym} \rangle$. If the Kelvin-Voigt model, where both quantities are equal, was applicable, all data points would collapse on the first bisectrix (black line). }}
\label{fig:kelvinvoigt}
\end{figure}

\modif{To perform an independent and more visual determination of $\tau$}, a simulation with a Maxwell model \modif{(Fig. \ref{fig:suppviscoelasticsimu})} qualitatively reproduces the cell deformations around the obstacle if we use the measured value  $\tau=70$~min\modif{, while} taking a much shorter \modif{value $\tau=10$~min} leads to almost no cell deformation around the obstacle \modif{and} taking a much longer \modif{value $\tau=700$~min} leads to cell deformations much higher than the one experimentally observed. A fortiori, our data rule out \modif{viscoelastic} solid behaviour models in which the elastic deformation could be sustained indefinitely (which is the limit of an infinite $\tau$). \modif{As a third independent test, Fig. \ref{fig:kelvinvoigt} rejects the predictions of the Kelvin-Voigt \modif{viscoelastic} solid model}\fg{, according to which the  time averaged cellular strain rate and  the tissue strain rate are equal; if this model was applicable, all data points would collapse on the first bisectrix (black line).}

\modif{This} \modif{estimation of $\tau$ could be compared to similar or related measurements from the literature, which for different cell types vary over orders of magnitude \cite{Vincent2015}, and specifically for MDCK monolayers range from 15~min \cite{Lee2011CrawlingSignaling} to $3 - 5$~h \cite{Vincent2015}. Our value falls within the range of one to a few hours required to explain the  onset of velocity \modif{waves} and strain waves  \cite{Tlili2018}. The relaxation time $\tau$ we find is associated with cell rearrangements} \modif{which fluidify the tissue by relaxing cell shapes \cite{Marmottant2009,Tlili2015}.}

Note that $\tau$ is much longer than the viscoelastic time associated with \modif{intracellular} stress dissipation due to cytoskeleton viscosity ($\approx$ 15 min \cite{etienne2740}) or cell shape relaxation time \modif{(of order of minute in Zebrafish tailbud \cite{Serwane2017} and even of second in a suspended MDCK monolayer \cite{Harris2012}). A tissue deformation is first transmitted to cell scale within a\fg{n} intracellular time scale, then rearrangements occur at intercellular time scale \cite{Phillips1978a,Marmottant2009}. It is the latter time \fg{that} is probed within the present set-up and found to be on the hour timescale.}

\modif{On the other hand, $\tau$ is much shorter than the viscoelastic time associated with cell division, typically several hours \cite{Ranft2010}. This \fg{suggests} that the cell division rate does not play a significant role in the monolayer fluidity, and changing it should not affect $\tau$. Such prediction is successfully tested in Fig. \ref{fig:superimpos}c and Fig. \ref{fig:suppdrugs}, where $\tau$ is insensitive to the mitomycin. 
In addition, $\tau$ is increased by myosin inhibitor blebbistatin, \fg{suggesting} \modif{that myosin activity contributes to fluidifying the tissue}. }
%\modif{The blebbistatin effect on $\tau$ is not an effect of 
%%%\modif{a decrease in cell-substrate interaction resulting in}
%the deformation rate being smaller, as this rate is also smaller in the  experiment without mitomycin (the migration velocity is lower in both cases). }

\begin{figure}[ht]
\begin{center} % a enlever si on veut utiliser l'environnement "figure"
\includegraphics[width=9cm]{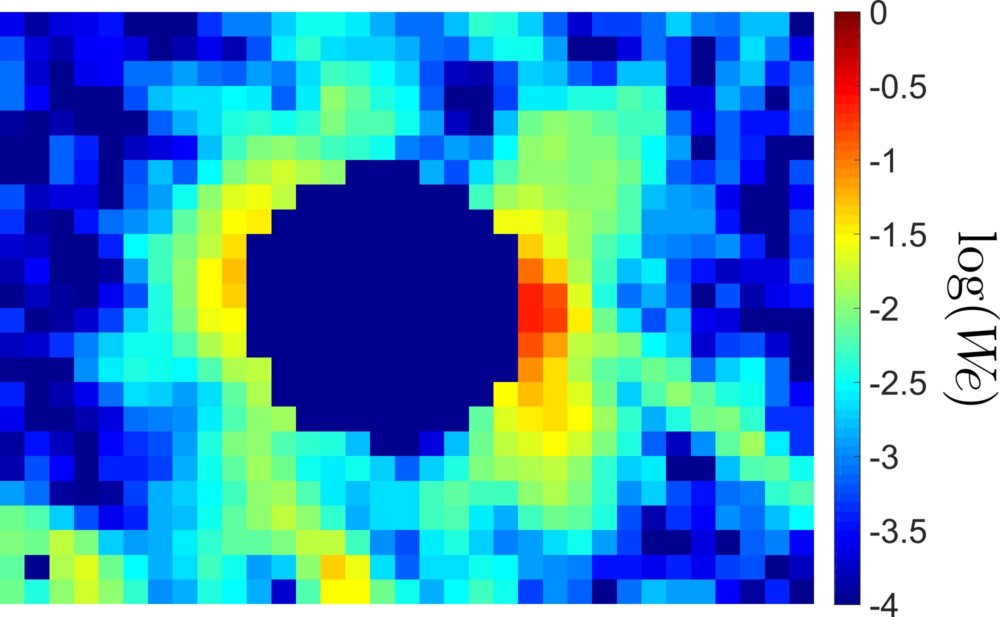}
\end{center} % a enlever si on veut utiliser l'environnement "figure"
\caption{\modif{Weissenberg number map. Same experiment and data as in  Fig.~\ref{fig:superimpos}a. One color per box, in decimal logarithm.}}
\label{fig:weissenberg}
\end{figure}

\modif{To interpret the influence of $\tau$ on the flow, it can be compared to the amplitude of the total deformation rate. The Weissenberg number $W\!e = \tau \vert \dot{\varepsilon}_{\rm tot}\vert $ is a dimensionless number characterizing how elasticity affects the flow and how reciprocally the flow affects cell shapes. In tissue regions where $W\!e \ll 1$ the flow is quasistatic, cell shapes relax and remain close to their rest
shape. Conversely, wherever  $W\!e$ is comparable \modif{to} or larger than 1, cell shapes depend on the tissue flow history. 
We find that $W\!e$ reaches its largest value $\approx 0.5$ just downstream of the obstacle (Fig. \ref{fig:weissenberg})\fg{. This value is at} the cross-over between these regimes. Elastic and viscous contributions to the flow are comparable. Hence, by regulating the migration velocity and/or the rearrangement dynamics, cells can biologically tune  $W\!e$ and the proportion of the elastic deformation they relax.}
\modif{This could be a way to switch progressively from a developing, fluid tissue to a mature, solid tissue \cite{Mongera2018}. Encoding in the cell shape the memory of the global tissue flow is a way to transmit information from tissue scale to cell scale, and can in turn influence intracellular signalling \cite{Jain2013CellGC}.}

\modif{The cell shape memory of the global tissue flow appears visually as the extension of the cell deformation wake, far downstream of the obstacle (see Fig. \ref{fig:kinematic}c), compared with the deformation rate causing this cell deformation which is much more localised near the obstacle \modif{(Fig. \ref{fig:kinematic}b)}.  
Cell deformation is advected downstream at a velocity of order of 1 $\mu$m/min (Fig. \ref{fig:kinematic}a) and its principal source of decay is rearrangements \fg{ since the total deformation rate essentially vanishes in this region}. This competition between deformation advection and rearrangements defines a characteristic length scale $\tau v$. This length scale is $\approx 100$~$\mu$m, i.e. it has the same order of magnitude  as the correlation length scale usually observed in a migrating MDCK cell monolayer \cite{Vedula2012}. 
}

%\paragraph{\textbf{Conclusion and perspectives}}

\modif{A tissue which solidifies during its maturation acquires a yield strain, validating the foam-tissue analogy \cite{Mongera2018}.
However, here, although the flow is visually similar to that of a soap froth \cite{Cheddadi2011}, we do not find such a yield strain\fg{. Indeed, the graph of $\varepsilon_{e}$ versus $\dot{\varepsilon}_{r}$
does not display any threshold behavior. I}n a developing tissue or in collective cell migration, neighbour changes do occur even at low applied deformation, and fluidify the material.
The velocity field results from cell activity, more precisely the interplay between collective migration in a band and boundary conditions imposed by the obstacle (as opposed to the one-dimensional migration velocity  in an obstacle free band, which only depends on local cell density  \cite{Tlili2018}). 
}

\modif{The Stokes flow geometry offers several advantages. It enables to establish a stationnary heterogeneous flow pattern, induce rearrangements, probe the monolayer rheology, discriminate between different models, and measure the viscoelastic time.
Our analysis method is based on the flow spatial heterogeneity, with a high enough Weissenberg number, to observe cell deformation, deformation rate and rearrangement  fields  with a variety of amplitudes and orientations, in several boxes. Flow stability in time (obtained here thanks to mitomycin) enables data averaging over a few successive images to improve the signal-to-noise ratio. Provided these requirements are met, our analysis is likely easier to apply to monolayers than other existing techniques to measure $\tau$ such as 
injected functionalized droplets \cite{Serwane2017,Mongera2018}
or 
tissue compression \cite{Forgacs1998,Schotz2008}. Measured fields are spatially smooth, their signal-to-noise ratio sufficient to calculate time and space derivatives. Despite the actual velocity fluctuations, it is possible to compute the transport terms of elastic deformation and integrate them along velocity field lines. 
This enables us to determine all fields involved in a continuum mechanics description, without requiring \modif{any} theoretical {\it a priori} \modif{assumption} on the velocity fields}.

\modif{To extend this method \textit{in vivo}, an experimental technique to introduce an obstacle in a Drosophila embryo has been recently developed:  laser-induced tissue cauterization burns a group of cells, attaching it to the vitelline membrane surrounding the embryo  \cite{RAUZI2017153}. Such mechanical perturbation can help unveil the underlying cause of the morphogenetic tissue flow.
In the same spirit, magnetic fluid drops could  be introduced as obstacles in 3D tissues such as in Zebrafish embryo \cite{Serwane2017,Mongera2018}. 
Finally, the present analysis method could be used to analyse a natural motion where a tissue flows around a small organ embedded in it \cite{manningarxiv}. }

\begin{acknowledgments}

\modif{We thank Sri Ram Krishna Vedula for preliminary experiments that inspired this work, } Ibrahim Cheddadi, Philippe Marcq and Pierre Saramito  for stimulating discussions. \modif{B.L. acknowledges financial supports from the European Research Council under the European Union's Seventh
Framework Programme (FP7/2007-2013) / ERC grant agreement n° 617233 and Agence Nationale de la Recherche (ANR) “POLCAM” (ANR-17-CE13-0013).}

\end{acknowledgments}

% \newpage

%\bibliography{bibliostockespaper,stokesh,stokes_mend,footnote}
\bibliography{bibliostockespaper}

\newpage

 \beginsupplement

\begin{center}  

\modif{\Large A migrating epithelial monolayer flows \fg{like} a Maxwell viscoelastic liquid}
\medskip \medskip

\modif{\Large by S. Tlili et al. }
\end{center} 

\medskip \medskip \medskip \medskip \medskip \medskip \medskip \medskip \medskip \medskip

{\Large \bf Supplementary Material}

\medskip \medskip \medskip \medskip \medskip \medskip \medskip \medskip \medskip \medskip

\section*{Calculation of transport and rotation terms}

\begin{figure}[ht]
\begin{center} % a enlever si on veut utiliser l'environnement "figure"
  \includegraphics[width=17.9cm]{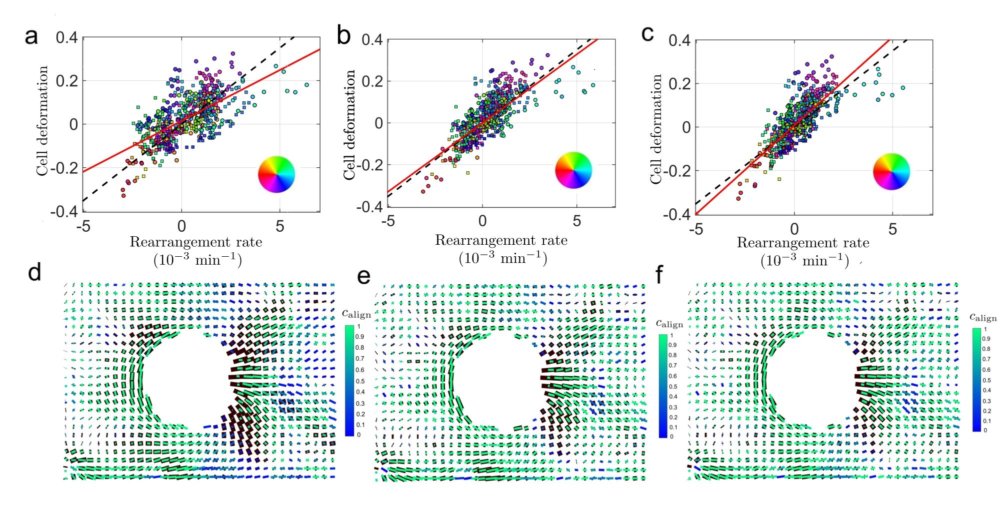}
\end{center}
\caption{\modif{Coarse-grained cell anisotropy} transport and rotation terms. Graphs \modif{of $\varepsilon_{\rm e}$ vs $\dot{\varepsilon}_{r}$} $(a-c)$  and maps \modif{of $\dot{\varepsilon}_{r}$} $(d-f)$, determined as $\dot{\varepsilon}_{r} = \dot{\varepsilon}_{tot}-\dot{\varepsilon}_{\rm e}= (\nabla v +\nabla v^{T})/2 - D\varepsilon_{\rm e}/Dt$, using different approximations to estimate $D\varepsilon_{\rm e}/Dt$. 
(a,d) Neglecting both transport and rotation terms, $D\varepsilon_{\rm e}/Dt \approx \partial \varepsilon_{\rm e}/\partial t$. 
(b,e) Taking into account transport but neglecting rotation, $D\varepsilon_{\rm e}/Dt \approx 
\partial \varepsilon_{\rm e}/\partial t + \vec{v}\cdot \nabla \varepsilon_{\rm e}$; same data as Fig.~\ref{fig:superimpos}b. (c,f) Complete expression taking into account both transport and rotation (see details in \modif{Supplementary Materials and in} \cite{Tlili2015}). 
Same  individual experiment as in  Figs. \ref{fig:exp_base}, \ref{fig:kinematic}; obstacle diameter 200~$\mu$m.
 In graphs \fg{(a-c)}, 
 %\modif{same representation as in Fig.~\ref{fig:superimpos}b: circles represent}
 $ \langle\left(\varepsilon_{\rm e}\right)_{xx}-\left(\varepsilon_{\rm e}\right\rangle_{yy})/2$  \modif{is} plotted vs  $ \langle\left(\dot{\varepsilon}_{r} \right)_{xx}-\left(\dot{\varepsilon}_{r}\right\rangle_{yy})/2$ while 
 % \modif{squares represent} 
 $\langle\left(\varepsilon_{\rm e}\right)_{xy}\rangle$
\modif{is} plotted vs $\langle\left(\dot{\varepsilon}_{r}\right)_{xy}\rangle$; 
\modif{data with $\varepsilon_{\rm e}^{\rm dev}$ amplitude smaller than 0.05 are excluded from the fit}.
\modif{Dashed b}lack lines\modif{: slope} $\tau = 70$~min\modif{. Solid r}ed lines\modif{: linear fit to the data, passing through the origin, slope} $\tau = 47$~min and $R = 0.62$ for (a), $\tau = 68$~min and $R = 0.73$ for (b)\modif{,} $\tau = 79$~min and $R = 0.73$ for (c). In \modif{each map, same representation as in Fig.~\ref{fig:superimpos}a, same $\tau$ value as in the above graph.}
}
\label{fig:supp2transport}
\end{figure}

\begin{figure}[ht]
\begin{center} % a enlever si on veut utiliser l'environnement "figure"
\includegraphics[width=17.9cm]{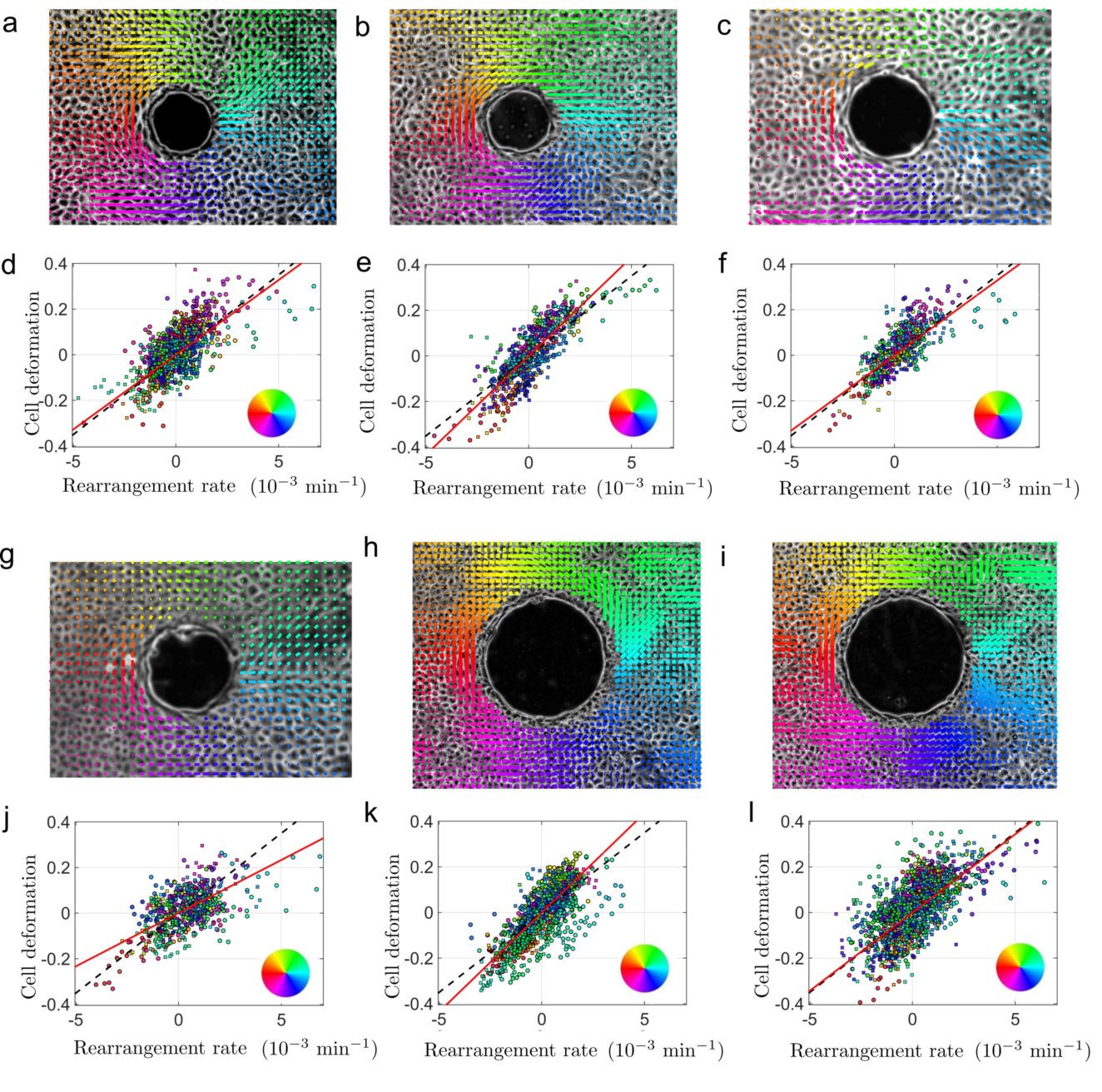}
\end{center} % a enlever si on veut utiliser l'environnement "figure"
\caption{Determination of viscoelastic time.
Three experiment dimensions, each being tested twice for reproducibility: $(a,b,d,e)$ obstacle diameter 150~$\mu$m, strip width 750~$\mu$m; $(c,f,g,j)$ obstacle diameter 200~$\mu$m, strip width 1000~$\mu$m; $(h,i,k,l)$  obstacle diameter 300~$\mu$m, strip width 1000~$\mu$m. 
Top panels (a,b,c,g,h,i): Deviatoric part of the cell deformation tensor; the positive extension axis is represented as a bar. \modif{The color codes for the  box angle position, in polar coordinate originating at the obstacle center, as in Fig. \ref{fig:kinematic}b,c. For legibility, only the cell deformation is represented, not the rearrangement rate.}
Bottom panels (d,e,f,j,k,l): Cell deformation vs rearrangement rate. Components of the deviatoric tensors are plotted for the six strips, with the same color code as in top panels.
\modif{Same representations as in Fig.~\ref{fig:supp2transport}.
\modif{Solid red line: linear fit to the data, passing through the origin; dashed black} line: slope 70~min. Note that (f) has the same data as Fig.~\ref{fig:superimpos}b. \modif{$R$ values: 0.65., 0.78, 0.73, 0.50, 0.76, 0.59.}
}
}
\label{fig:supp3obstacles}
\end{figure}

\begin{figure}[ht]
\begin{center} % a enlever si on veut utiliser l'environnement "figure"
\includegraphics[width=8cm]{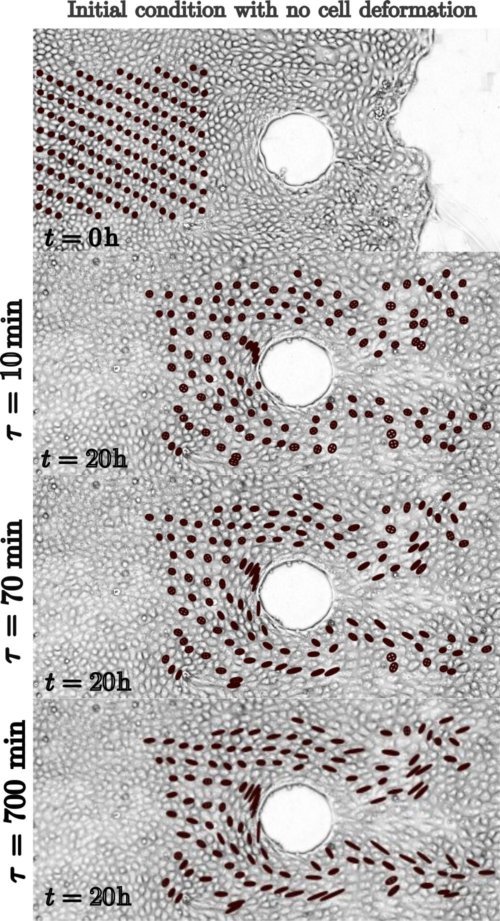}
\end{center} % a enlever si on veut utiliser l'environnement "figure"
\caption{\modif{Visual test of the viscoelastic time value. Assuming the tissue is a viscoelastic liquid with a characteristic time $\tau$, we simulate Lagrangian trajectories and deformations of ``virtual" cells, which are simply tracers placed on the tissue. $(a)$ Initial condition: 200 tracers placed on a square 40 $\mu$m spaced grid. A tracer labeled $i$ is assigned a size equal to the actual experimental average cell size in this box, and an initially null deformation $\varepsilon_{\rm e}^i(0) = 0$. The time step is $dt = 1$ frame $= 5$~min. 
At each time-step, the tracer position $\vec{r}_{i}(t)$ is advected with its actual experimental velocity field according to  $\vec{r}_{i}(t+dt) = \vec{r}_{i}(t) + dt\cdot\vec{v}(\vec{r}_{i}(t),t)$. \modif{The deviatoric part of its} deformation is updated according to \modif{a viscoelastic behaviour,} $\varepsilon_{\rm e}^{i}(t+dt)  = \varepsilon_{\rm e}^{i}(t)+dt\cdot \left(\left[\nabla v(\vec{r}_{i}(t),t) +\nabla v^{T}(\vec{r}_{i}(t),t)\right]/2-\varepsilon_{\rm e}^{i}(t)/\tau \right)$ \modif{while the trace of its deformation is updated according to an elastic behaviour}. $(b,c,d)$ Tracer deformation calculated after 24 hours, for three different viscoelastic time values. $(b)$ With $\tau = 10$~min, the deformation is quickly relaxing as for a pure \modif{liquid behavior}, and remains much lower than in experiment. $(c)$ With $\tau = 70$~min, which is the value found in experiments (Fig. \ref{fig:superimpos}a,b, Fig. \ref{fig:supp3obstacles}), elongations are similar to the experimentally observed elongations, and both elastic and \modif{viscous} contributions are simultaneously \modif{significant}.
 $(d)$ With $\tau = 700$~min, deformations are much higher than in experiments, as in a pure elastic behavior.  }}
\label{fig:suppviscoelasticsimu}
\end{figure}

\modif{Complete transport and rotation terms are calculated according to Eq. 20 of Tlili et al. \cite{Tlili2015}, and used to plot $\dot{\varepsilon}_{r}$ in Fig.~\ref{fig:supp2transport}c,f. In brief, the complete  evolution equation for $\varepsilon_{\rm e}$ writes:
$$
    \partial_{\rm t} \varepsilon_{\rm e} + v\nabla \varepsilon_{\rm e}= \frac{\nabla v+\nabla v ^{T}}{2}-\dot{\varepsilon}_{r}-\frac{\alpha_{\rm g}I}{2} + \left(\nabla v -\dot{\varepsilon}_{r} -\frac{\alpha_{\rm g}I}{2}\right)\varepsilon_{\rm e} + \varepsilon_{\rm e}\left(\nabla v^{T} -\dot{\varepsilon}_{r} -\frac{\alpha_{\rm g}I}{2}\right)
$$
where $I$ is the unit tensor, $\alpha_{\rm g} = \mathrm{Tr}(\nabla v)$, and for off-diagonal ($xy$) components we use the convention:
\begin{itemize}
    \item  $\vec{v} = [u,v]$,
\item $\varepsilon_{\rm e}=
  \begin{bmatrix}
    e_{\rm xx} & e_{\rm xy}  \\
    e_{\rm xy} & e_{\rm yy} 
  \end{bmatrix}$,
 \item   $\dot{\varepsilon}_{r}=
  \begin{bmatrix}
   d_{\rm xx} & d_{\rm xy}  \\
    d_{\rm xy} & d_{\rm yy} 
  \end{bmatrix}$,
  \item $
 \nabla v=
  \begin{bmatrix}
   u_{\rm x} & u_{\rm y}  \\
    v_{\rm x} & v_{\rm y} 
  \end{bmatrix}$,
\item  
$ \nabla v^{T}=
  \begin{bmatrix}
   u_{\rm x} & v_{\rm x}  \\
    u_{\rm y} & v_{\rm y} 
  \end{bmatrix}
$.
\end{itemize}
Isolating  $\dot{\varepsilon}_{r}$ from the evolution equation, we find that we can compute it according to: $$\dot{\varepsilon}_{r} =  E^{-1} A$$ where we define:
\begin{itemize}
    \item $E = \begin{bmatrix}
    1+2e_{\rm xx} &0 & 2e_{\rm xy}\\
    0 &  1+2e_{\rm yy}  & 2e_{\rm xy}\\
    e_{\rm xy} & e_{\rm xy}  & 1+e_{\rm xx}+e_{\rm yy}
  \end{bmatrix} $,
     \item  $ A = \begin{bmatrix}
    A_{\rm xx}\\
    A_{\rm yy}\\
    A_{\rm xy}
  \end{bmatrix}$,
\end{itemize}
where the components of $A$ are:
\begin{itemize}
 \item $A_{\rm xx} = (1+2e_{\rm xx})d_{\rm xx}+2e_{\rm xy}d_{\rm xy} $ \\ $ ~ \quad \quad = u_{\rm x}+2u_{\rm x}e_{\rm xx}+2u_{\rm y}e_{\rm xy}-\alpha_{\rm g}e_{\rm xx}-\frac{\alpha_{\rm g}}{2}-(\partial_{\rm t} e_{\rm xx} + u\partial_{\rm x} e_{\rm xx}+ v\partial_{\rm y} e_{\rm xx})$,
 \item $A_{\rm yy} = (1+2e_{\rm yy})d_{\rm yy}+2e_{\rm xy}d_{\rm xy} $ \\ $ ~ \quad \quad  = v_{\rm y}+2v_{\rm y}e_{\rm yy}+2v_{\rm x}e_{\rm xy}-\alpha_{\rm g}e_{\rm yy}-\frac{\alpha_{\rm g}}{2}-(\partial_{\rm t} e_{\rm yy} + u\partial_{\rm x} e_{\rm yy}+ v\partial_{\rm y} e_{\rm yy})$,
 \item $A_{\rm xy} = (1+e_{\rm xx}+e_{\rm yy})d_{\rm xy}+e_{\rm xy}d_{\rm xx}+e_{\rm xy}d_{\rm yy}$ \\ $ ~ \quad \quad  = \frac{1}{2}(u_{\rm y}+v_{\rm x})+u_{\rm x}e_{\rm xy}+v_{\rm y}e_{\rm xy}+u_{\rm y}e_{\rm yy}+v_{\rm x}e_{\rm xx}-\alpha_{\rm g}e_{\rm xy}-(\partial_{\rm t} e_{\rm xy} + u\partial_{\rm x} e_{\rm xy}+ v\partial_{\rm y} e_{\rm xy}).$
\end{itemize}}

\newpage

\section*{Supplementary Movies}

\medskip \medskip \medskip \medskip \medskip \medskip \medskip \medskip \medskip \medskip

\begin{center}  
\includegraphics[height=4cm]{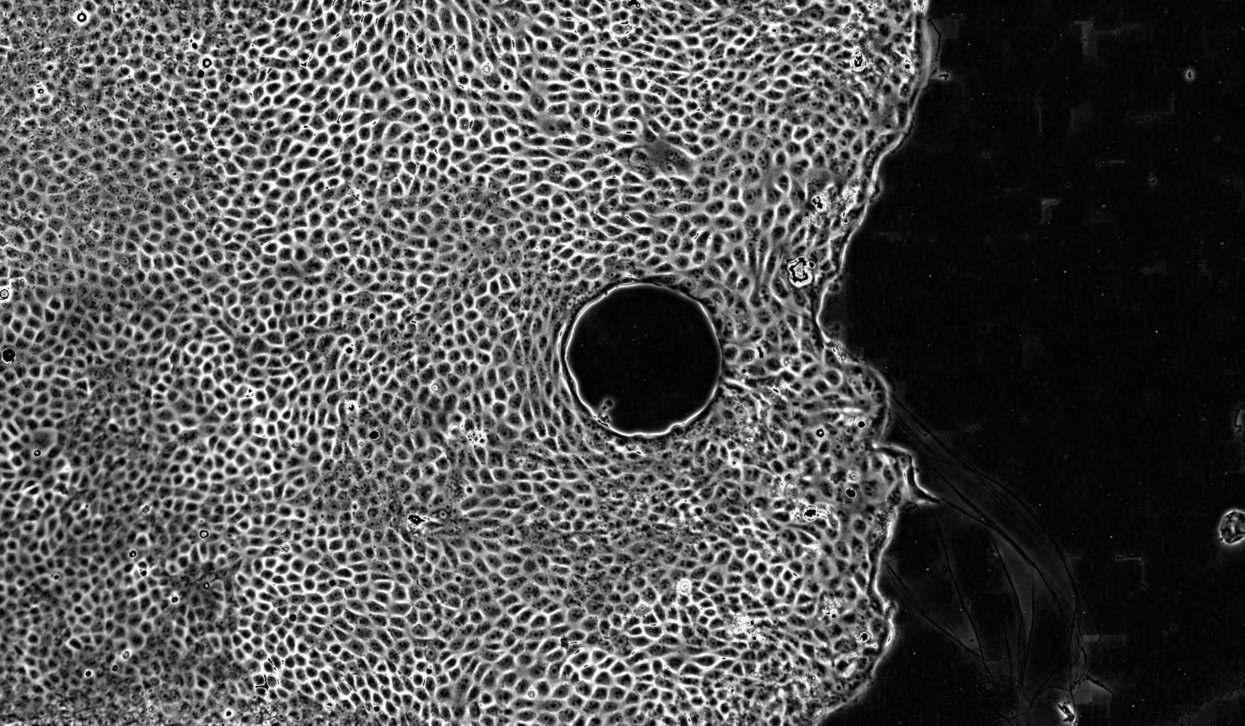}
\end{center} 
\modif{Supplementary Movie 1. Reference experiment, same as Fig.~\ref{fig:exp_base}; obstacle diameter 200~$\mu$m, strip width 1000~$\mu$m\modif{, movie and analysis durations  20~h}. Original version: 5~min time interval, pixel size 0.65~$\mu$m; low-size version: 15~min time interval, pixel size 2.6~$\mu$m.}

\medskip \medskip \medskip \medskip \medskip \medskip \medskip \medskip \medskip \medskip

\begin{figure}[ht]
\begin{center} % a enlever si on veut utiliser l'environnement "figure"
\includegraphics[width=12cm]{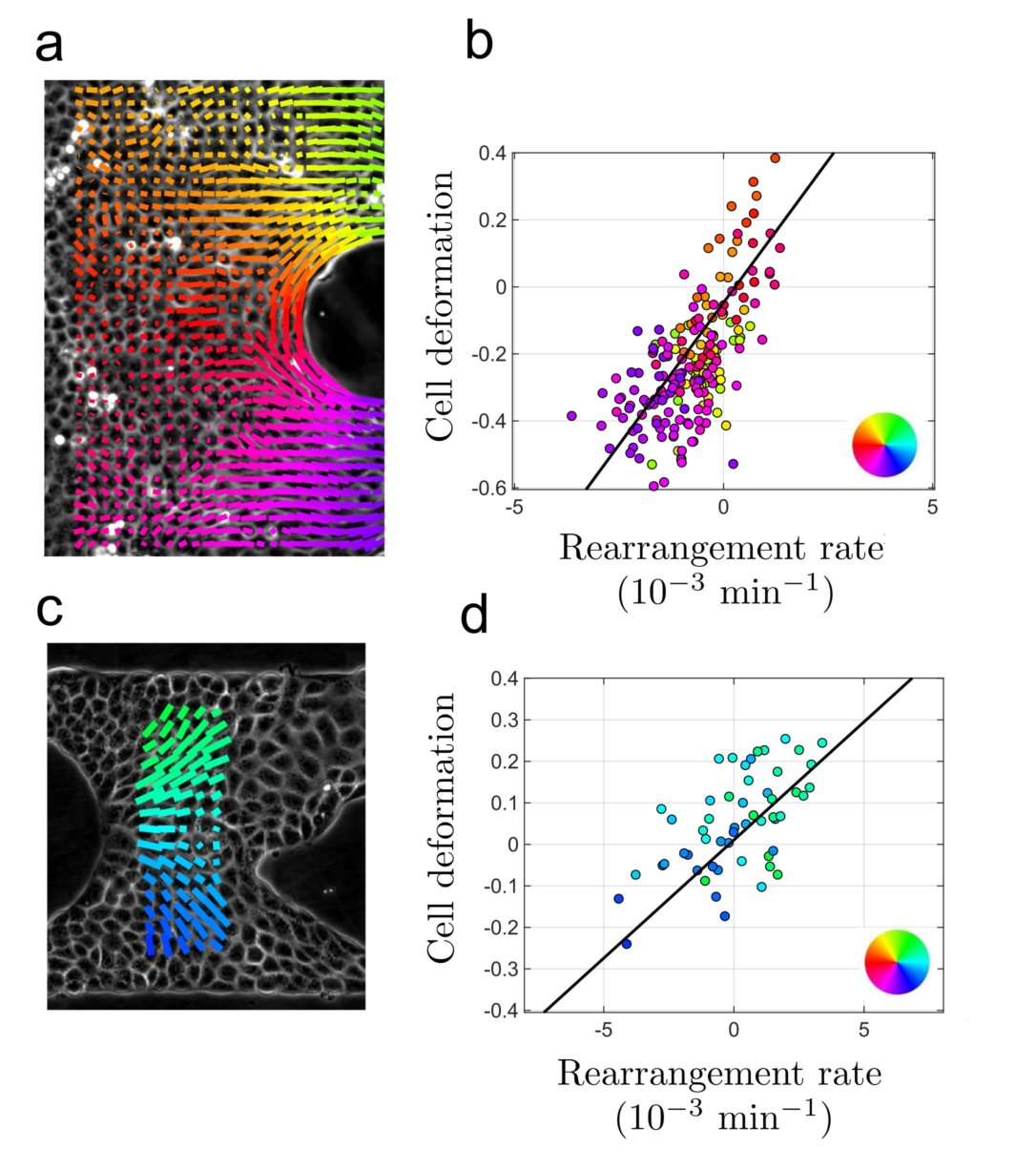}
\end{center} % a enlever si on veut utiliser l'environnement "figure"
\caption{\modif{Effect of drugs on viscoelastic time.
$(a,b)$ The myosin activity inhibitor blebbistatin (10~$\mu$M) is added to the medium at the beginning of the experiment; obstacle diameter 330~$\mu$m, strip width 1000~$\mu$m. See Supplementary Movie 2.
$(c,d)$ Division inhibitor mitomycin is {\it not} added; obstacle diameter 200~$\mu$m, strip width 300~$\mu$m. Due to cell density increase and jamming, cells migrate slowly. We analyze the region downstream of the obstacle, where in a few boxes significant cell deformation and migration are consistently observed during a period of 6 hours. See Supplementary Movie 3. 
In both experiments, \modif{since migration is impaired, only part of the boxes display large enough cell deformation and velocity, during a time long enough for averaging purpose, and can be used for the analysis}.  In the blebbistatin case, steady migration is only observed near the migrating front. In the case without mitomycin, divisions lead rapidly to density increase and tissue jamming at the back of the obstacle.}
}
\label{fig:suppdrugs}
\end{figure}

\begin{center}  
\includegraphics[height=4cm]{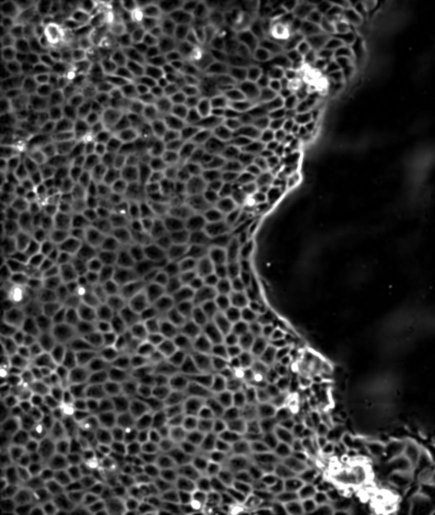}
\end{center} 
\modif{Supplementary Movie 2. Experiment with myosin activity inhibition, same as Fig.~\ref{fig:suppdrugs}a,b; obstacle diameter 200~$\mu$m, strip width 1000~$\mu$m, 5~min time interval\modif{, movie duration 12~h, analysis duration 6~h}. Original version: pixel size 0.65~$\mu$m; low-size version: pixel size 2.6~$\mu$m.}

\medskip \medskip \medskip \medskip \medskip \medskip \medskip \medskip \medskip \medskip
 
\begin{center}  
\includegraphics[height=4cm]{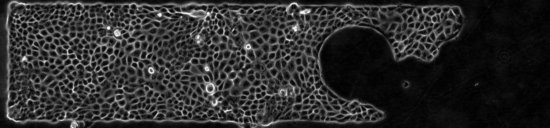}
\end{center} 
\modif{Supplementary Movie 3. Experiment without division inhibition, same as Fig.~\ref{fig:suppdrugs}c,d; obstacle diameter 200~$\mu$m, strip width 300~$\mu$m, 5~min time interval\modif{, movie duration 12~h, analysis duration 4~h}. Original version: pixel size 0.65~$\mu$m; low-size version: pixel size 2.6~$\mu$m.}

\newpage

\section*{Supplementary Figures}

\medskip
\medskip
\medskip
\medskip

\newpage

\newpage
% $ \langle\left(\varepsilon_{\rm e}\right)_{xx}-\left(\varepsilon_{\rm e}\right)_{yy}\rangle/2$
% is plotted vs  $ \langle\left(\dot{\varepsilon}_{r} \right)_{xx}-\left(\dot{\varepsilon}_{r}\right)_{yy}\rangle/2$
% and $\langle\left(\varepsilon_{\rm e}\right)_{xy}\rangle$
% is plotted vs $\langle\left(\dot{\varepsilon}_{r}\right)_{xy}\rangle$.
% Data for deviator of $\varepsilon_{\rm e}$ with an amplitude smaller than 0.05 are not included in the fits.
% In maps, the color codes for the \modif{box angle} position, in polar \modif{coordinate originating at} the obstacle center, as in Fig. \ref{fig:kinematic}b,c.

\newpage

\clearpage

\newpage

\newpage

\newpage

\clearpage

\end{document}